\documentclass[prb,preprint]{revtex4-1} 

\usepackage{amsmath}
\usepackage{amsfonts}
\usepackage{graphicx,epstopdf}
\usepackage{grffile}
\usepackage{color}

\newcommand\myeq{\mathrel{\overset{\makebox[0pt]{\mbox{\normalfont\tiny\sffamily def}}}{=}}}

\begin{document}

\title{Gravity Tunnel Drag}

\author{Thomas G. Concannon}
\email{thomasconcannon@kings.edu}
\author{Gerardo Giordano}
\email{gerardogiordano@kings.edu}
\affiliation{Department of Chemisty and Physics, King's College, Wilkes-Barre, PA 18711}

\date{\today}

\begin{abstract}
The time it takes to fall down a tunnel through the center of the Earth to the other side takes approximately 42 minutes, but only when given several simplifying assumptions: a uniform density Earth; a gravitational field that varies linearly with radial position; a non-rotating Earth; a tunnel evacuated of air; and zero friction along the sides of the tunnel. Though several papers have singularly relaxed the first three assumptions, in this paper we relax the final two assumptions and analyze the motion of a body experiencing these types of drag forces in the tunnel. Under such drag forces, we calculate the motion of a transport vehicle through a tunnel of the Earth under uniform density, under constant gravitational acceleration, and finally under the more realistic Preliminary Reference Earth Model (PREM) density data. We find the density profile corresponding to a constant gravitational acceleration better models the motion through the tunnel compared to the PREM density profile, and the uniform density model fares worse.

\end{abstract}

\maketitle 

\section{Introduction} 

It has long been the goal of humanity to minimize travel time between two places, be they on Earth or anywhere else for that matter. Technological advances in the past century or so have significantly cut down on travel time between two points on the Earth (e.g., the invention of the airplane). Some more imaginative people have envisioned drilling a tunnel through the Earth to get to the other side, and letting gravity do the work. This so-called ``gravity train" has a facinating history\cite{maa_simoson}, and the physics of it has been analyzed many times\cite{romer}. In fact, it is now a common introductory physics problem\cite{halliday}.

So let us reproduce here the usual calculation to find the free fall time through the tunnel, i.e., with several simplifying assumptions. First, assume the engineering project to build the tunnel through the center of the Earth was completed, as was the task of evacuating the tunnel of air. Second, assume that the transport vehicle did not experience any significant friction along the side walls of the tunnel (e.g., perhaps the support rails were superconductors and the transport vehicle levitated above them). Third, assume the Earth has a constant density throughout its interior and it has a gravitational field that varies linearly with radial position. Finally, assume the Earth is spherical and non-rotating so as to not put undue stress from gravitational torque on the workers. 

\begin{figure}[h!]
\includegraphics[width=50mm]{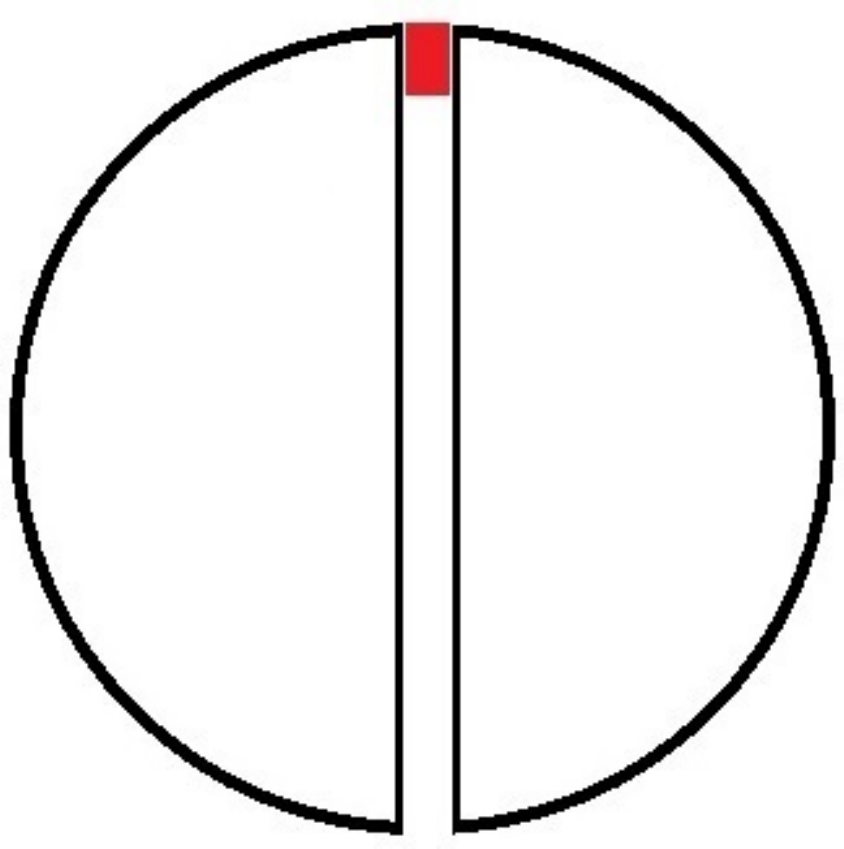}
\caption{Tunnel through the Earth.}
\label{earth_tunnel}
\end{figure}

With these assumptions, the free fall motion through the tunnel can easily be found by application of Newton's Shell Theorem. Let $\rho_0$ be the constant density of the Earth so that the mass of the Earth may be written 
\begin{equation}
M=\frac{4}{3}\pi R^3 \rho_0. \label{M_const_den_nofric} 
\end{equation}
where $R$ is the average radius of the Earth. Let the mass of the falling body be $m$. When this mass is a distance $r$ from the center of Earth, it feels a gravitational force arising from only the amount of mass of Earth inside radius $r$. In other words,
\begin{equation}
F_r=-\frac{GmM(r)}{r^2}=-\frac{Gm(4\pi r^3\rho_0/3)}{r^2}=-\frac{4\pi Gm\rho_0}{3}r=m\ddot{r} \label{F_const_den_nofric}
\end{equation}
where $M(r)$ is the mass of Earth within a sphere of radius $r$. Thus the motion is described by the second order
differential equation
\begin{equation}
\ddot{r}+\frac{4\pi G\rho_0}{3}r=0, \label{osc_nofric}
\end{equation}
which is immediately recognized as simple harmonic motion with an angular frequency of
$\omega^2=4\pi G\rho_0/3$ and a subsequent period of
\begin{equation}
T=\frac{2\pi}{\omega}=\sqrt{\frac{3\pi}{G\rho_0}}. \label{T_const_den_nofric}
\end{equation}
Since the gravitational acceleration is given by
\begin{equation}
g=\frac{GM}{R^2}=\frac{4}{3}\pi G\rho_0 R, \label{g_const_den_nofric}
\end{equation}
where $R$ is the average radius of the (spherical) Earth, the period of oscillation may also be 
written as
\begin{equation}
T=2\pi \sqrt{\frac{R}{g}} \approx 84\; \text{min}. \label{period84min}
\end{equation}

Given these assumptions, the travel time from one end of the tunnel to the other via free fall is thus roughly 42 minutes. 

Beyond this overly idealistic situation, however, other authors examined the motion of an object traveling through the tunnel by relaxing one of the assumptions at a time (for a few examples, see here\cite{maa_simoson} and here\cite{klotz}). Our aim in this paper is to relax the assumption of frictionless travel through the tunnel. First, we examine the case of travel through a non-evacuated tunnel in the specific case of quadratic drag and,  given the extreme pressures and densities as depth increases, we show that this is a practical impossibility even when the drag coefficient is unrealistically small. Second, we calculate how much friction would be needed to make the tunnel travel cost-effective compared to a meridional airplane trip, which will give us the frictional force (per unit mass) we will use throughout the rest of the paper. Having ruled out the practicality of motion through a non-evacuated tunnel, we then consider travel through an evacuated tunnel but such that the transport vehicle experiences constant sidewall friction (e.g., along a set of rails). For this case, we analyze the motion under three different density profiles of the Earth: uniform density, a recent model by Klotz\cite{klotz} which results in a constant gravitational acceleration through the Earth, and, finally, to make the result more realistic, the Preliminary Reference Earth Model (PREM) density data, which reconstructs the internal structure of the Earth using seismic data.

\section{Frictional Forces}

\subsection{Air Friction Through the Tunnel}
\begin{figure}[h!]
\includegraphics[width=140mm]{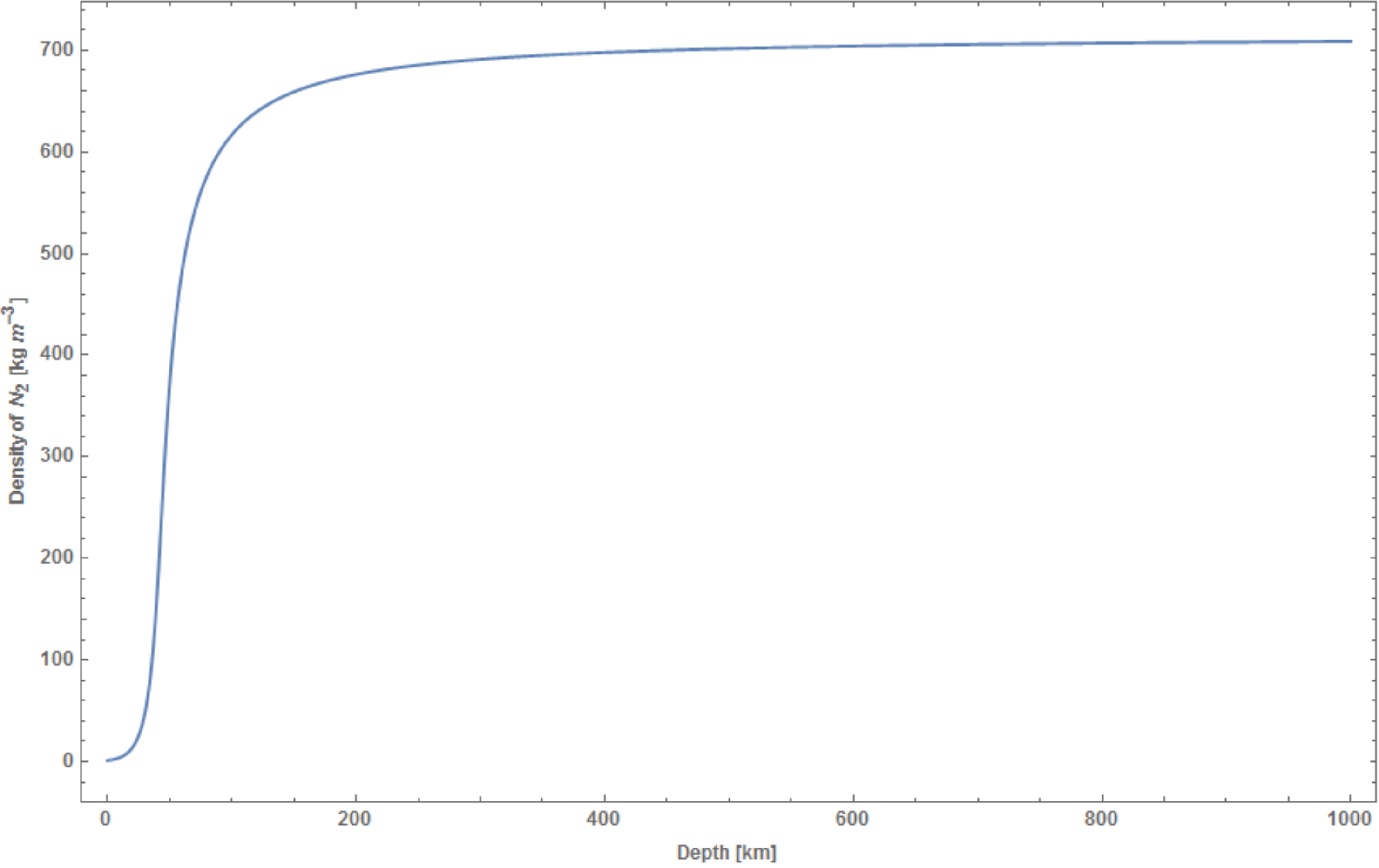}
\caption{Density as a function of depth within the Earth (using typical van der Waal parameters $a$ and $b$ found in Berberan-Santos, \emph{et al.} \cite{barometric}).}
\label{Density_as_function_of_depth}
\end{figure}

In most everyday situations humans experience, when the pressure and density of the surrounding medium approximates those at the surface of the Earth and before the onset of turbulence, the frictional drag force on a falling object can be modeled such that it is proportional to the square of the instantaneous velocity. The quadratic drag force is found from 
\begin{equation}
F=\frac{1}{2}\rho\,CAv^2, \label{quad_fric}
\end{equation} 
where $C$ is the drag coefficient, $\rho$ the density of the fluid, and $A$ the cross-sectional area of the object moving through the fluid. It can also be modeled such that the drag force is directly proportional to the instantaneous velocity itself, linear drag, described simply by $F=Cv$, but for more realistic motion, we will only consider a quadratic drag force in the following analysis. In the situation of the air in the tunnel through the Earth, the pressure and density within the Earth quickly rises to extreme values as a function of depth, so that the air in the tunnel cannot be simplistically modelled as an ideal gas\cite{barometric}. Modeling the air as a more realisitic van der Waal gas also reveals that the density of the air increases dramatically with depth, but so much so as to severely stifle the motion of a falling body such that it would never get much beyond the center of the Earth and would exhibit damped harmonic motion about the center with a maximum amplitude of only a few tens of kilometers. The barometric equation for a van der Waals gas is given by \cite{barometric}
\begin{equation}
\frac{d\rho}{dr}=-\frac{g(r)M\rho}{\frac{R_gTM^2}{(M-b\rho)^2}-\frac{2a\rho}{M}}, \label{barometric_formula}
\end{equation}
where $M$ is the molar mass of the gas within the tunnel, $R_g$ the gas constant, $a$ and $b$ the van der Waals parameters, $\rho$ the density of the air within the tunnel, and $g(r)$ is the acceleration due to gravity as a function of position (depth). If we assume the gas within the tunnel is nitrogen for simplicity, the gravitational acceleration profile $g(r)$ is found assuming a uniform density of the Earth, and allow Mathematica to solve equation \eqref{barometric_formula} for the density $\rho$ of the air in the tunnel as a function of depth within the Earth, graphing it reveals just how quickly the density of the air rises, nominally prohibiting realistic travel beyond a depth of 175 km, much less any kind of harmonic motion about the center of the Earth (Figure \ref{Density_as_function_of_depth}).

\begin{figure}[h!]
\includegraphics[width=140mm]{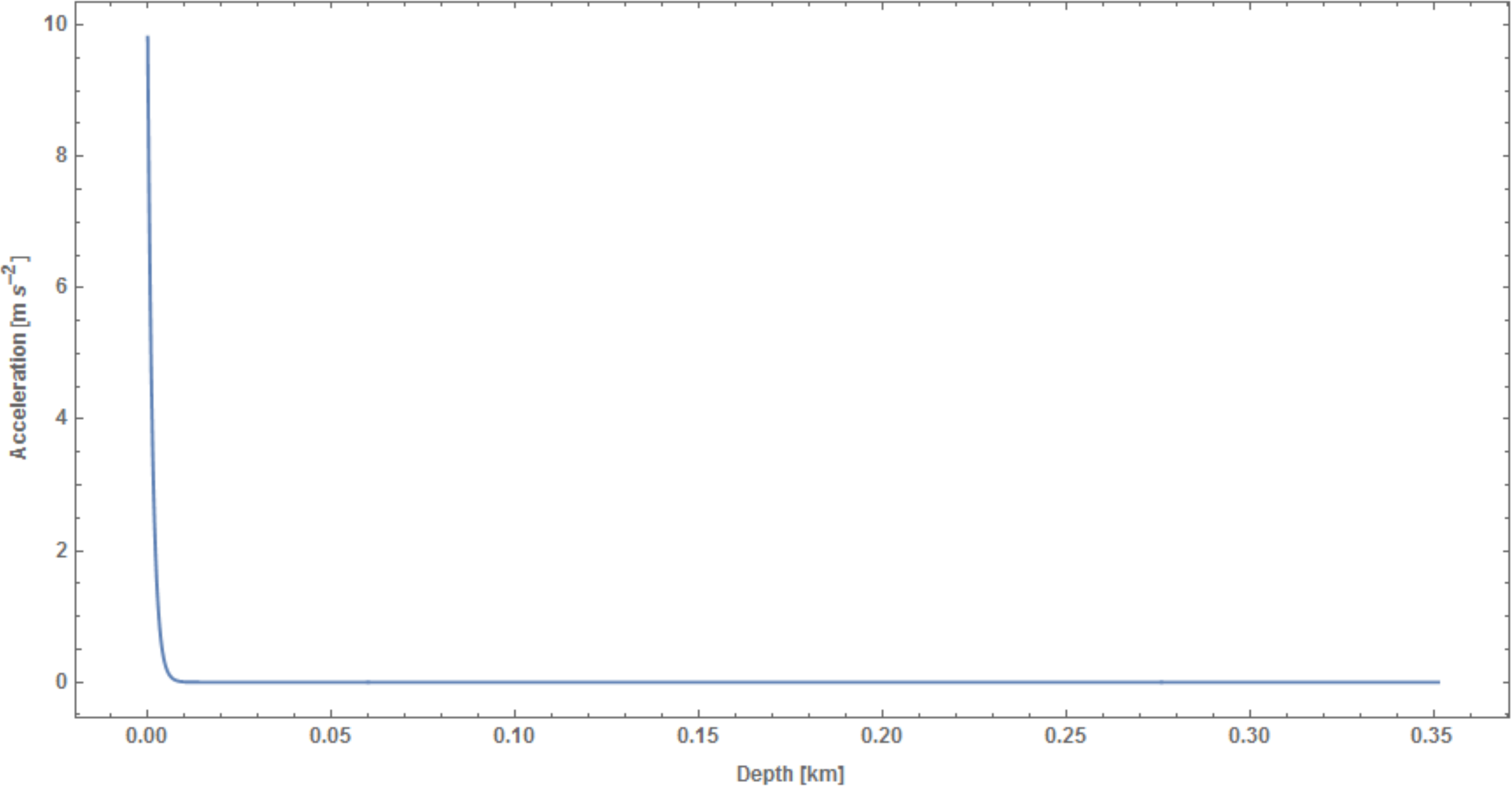}
\caption{Acceleration as a function of depth ($C \approx 0.015$).}
\label{AccVsPos3}
\end{figure}

\begin{figure}[h!]
\includegraphics[width=140mm]{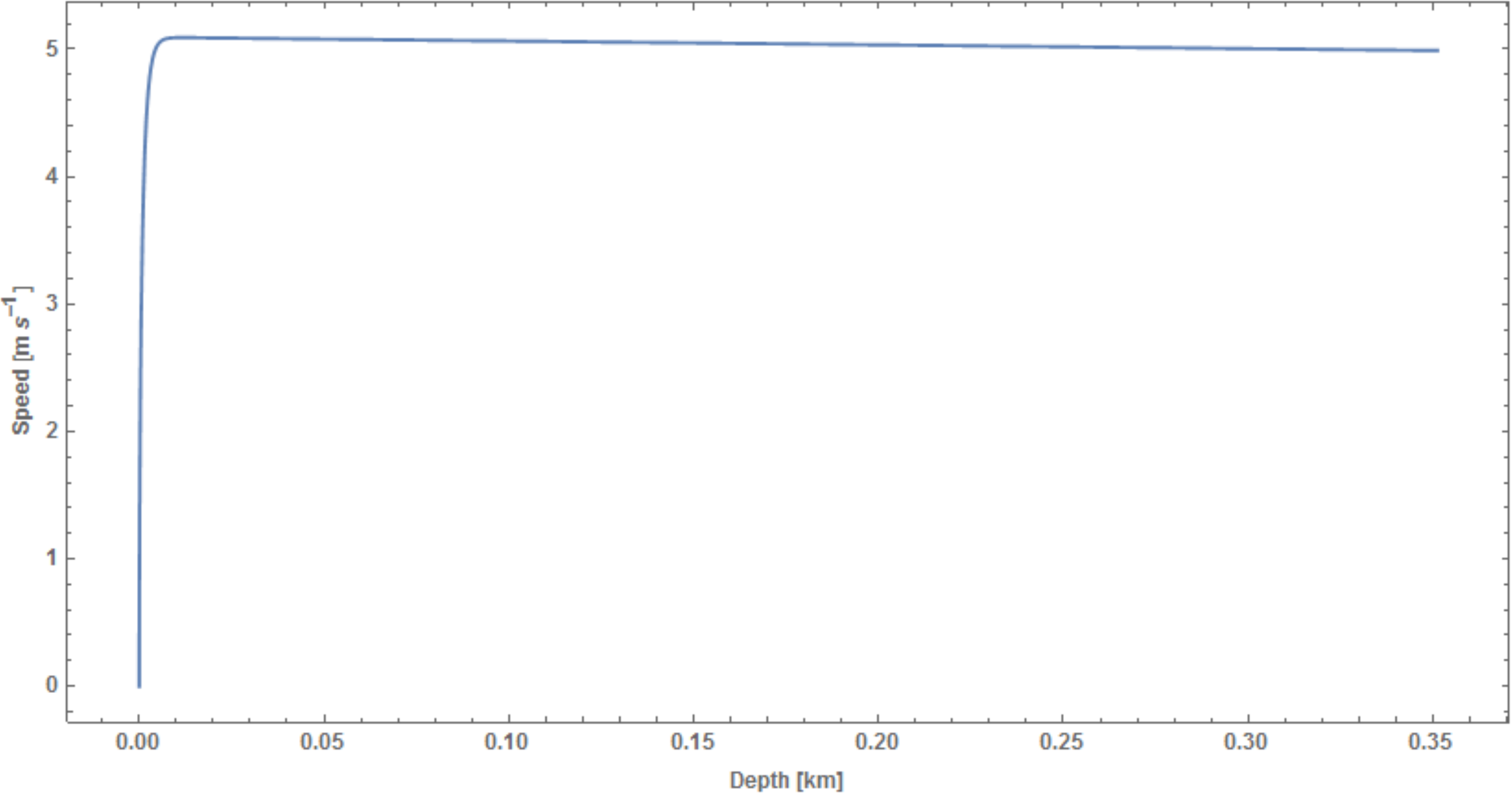}
\caption{Speed as a function of depth ($C \approx 0.015$).}
\label{SpeedVsPos3}
\end{figure}

\begin{figure}[h!]
\includegraphics[width=140mm]{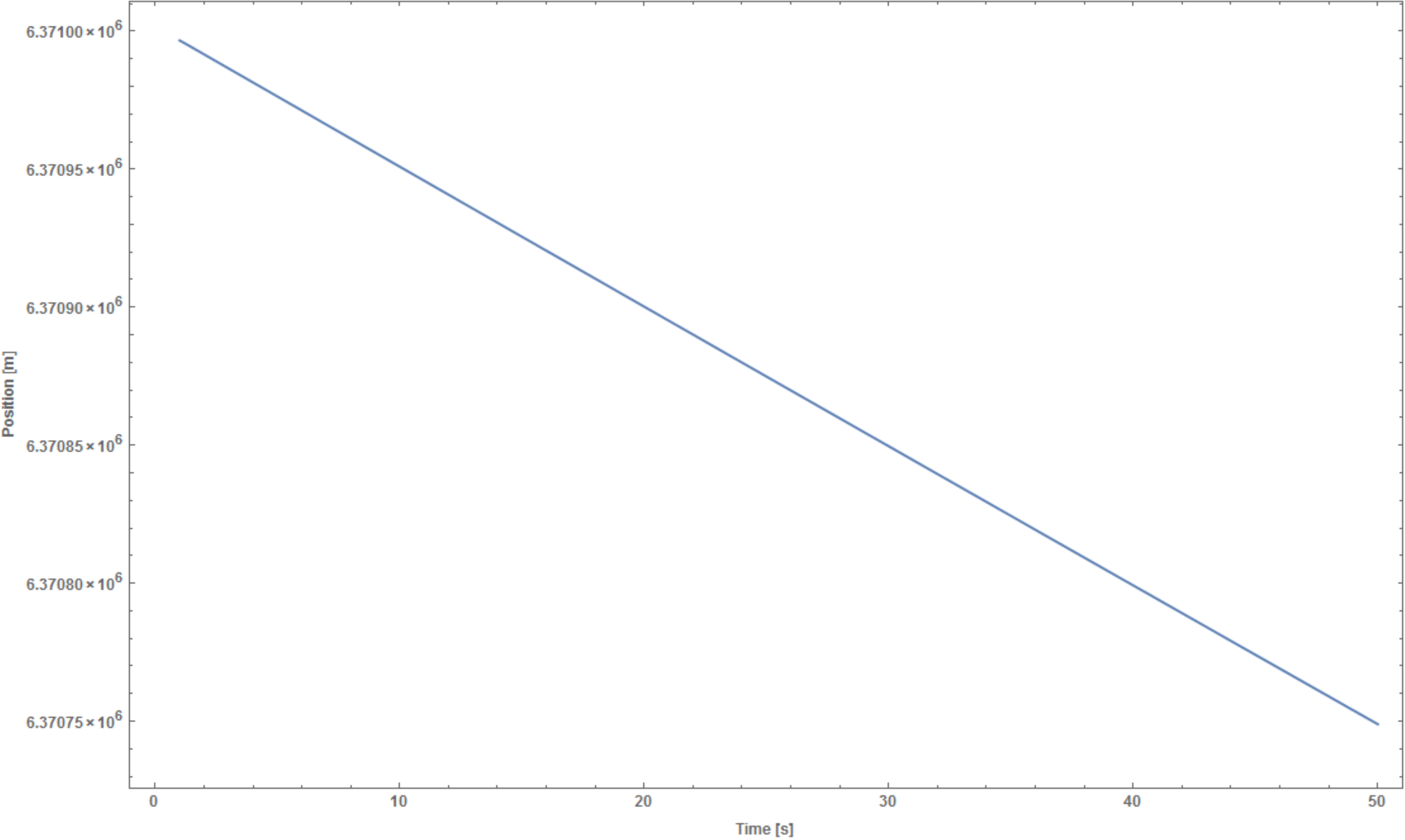}
\caption{Position as a function of time ($C \approx 0.015$).}
\label{DepthVsTime3}
\end{figure}

Finding the acceleration as a function of depth using equations \eqref{quad_fric} and \eqref{barometric_formula} dramatically shows how quadratic friction effects the motion. If we use realistic constants such that the cross-sectional area and drag coefficient correspond to that of an Airbus A380 \cite{airbus}, the transport vehicle reaches terminal velocity almost immediately (Figure \ref{AccVsPos3}). To picture the motion more easily, we include the graphs of the speed as function of depth and the depth of the vehicle as a function of time (Figures \ref{SpeedVsPos3} and \ref{DepthVsTime3}).

\begin{figure}[h!]
\includegraphics[width=140mm]{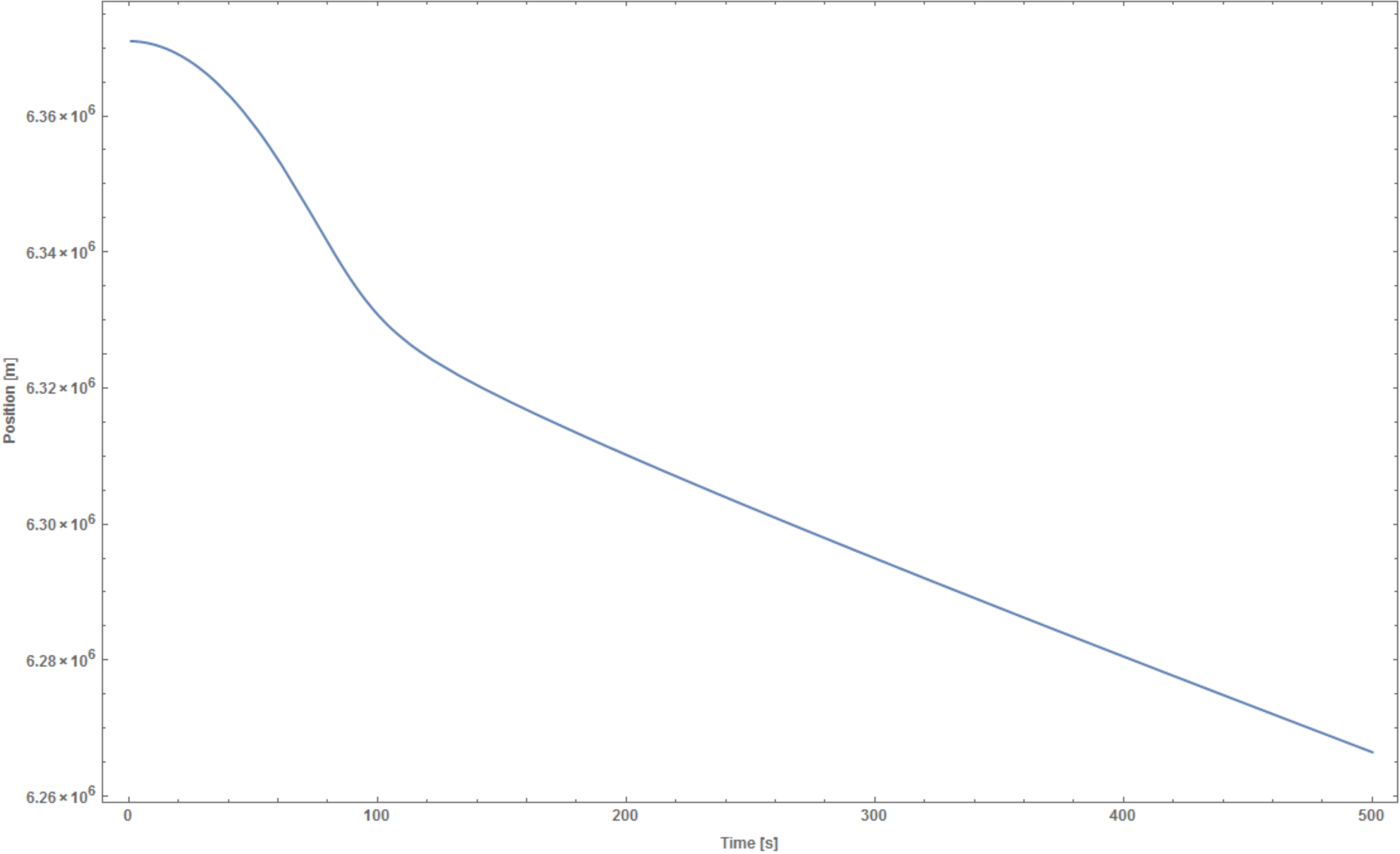}
\caption{Position as a function of time ($C \approx 4 \times 10^{-8}$).}
\label{DepthVsTime008}
\end{figure}

As a result of the above analysis, we conclude that it is practically impossible to consider a tunnel through the Earth that is not completely evacuated. The more realistic quadratic drag forces dramatically impede the motion; one can show that it takes $5.9 \times 10^7$ seconds, or approximately 1.8 years, to reach the center of the Earth, thus making it infeasible as a transport mechanism. Indeed, if technology made it possible to decrease the drag coefficient by six orders of magnitude, the quadratic drag forces would still impede the motion considerably, as Figure \ref{DepthVsTime008} clearly shows, and makes the trip to the Earth's center approximately 28 hours. Much worse, however, is the fact that the air deep within the tunnel cannot be treated as a van der Waals gas because nitrogen gas will solidify around 420 km down under the tremendous pressure \cite{liquid_vapor}. Henceforth, and quite reasonably so, we must consider motion through an evacuated tunnel. 

\subsection{Energy Considerations}
To keep the gravity tunnel competitive with other modes of transportation, for example, an airplane, would require that the energy necessary to bring the transport back to the surface from its resting place is less than the energy used by the airplane. As an exercise, let us consider the Airbus A380 with a loaded mass of 575,000 kg, an operating speed of 945 km/h, and an average fuel efficiency of approximately 0.0475 km/L for the flight while loaded with fuel and passengers\cite{airbus2}. To travel half of the meridional circumference, a journey of about 20,000 km would require at least 21 hours of travel time (not to mention stops for refueling) and approximately 500,000 liters of jet fuel or about $17 \times 10^{12}$ J of energy. As long as the energy losses due to friction in the gravity tunnel were less than $17 \times 10^{12}$ J, the gravity train would be more efficient than an airplane, and shrink a trip of about 21 hours to 42 mins. The energy lost due to friction would need to be added back to the train in order to avoid getting stuck well below the surface of the Earth. If we consider that energy to be the work done by friction, we can assume a frictional force of about 1,300,000 N (the equivalent of 4 jet engines) or about 2.33 N per kg of the transport vehicle. Using this value in the calculations that follow resulted in uniteresting results (i.e., the transport vehicle's motion is severely overdamped and quickly comes to rest at the core), so the value was lowered arbitrarily to about 15\% of this value, roughly 0.35 N per kg of the transport vehicle, in order to reveal some of the more interesting aspects of the resulting oscillatory motion.

\section{Tunnel With Constant Contact Friction}

As a result of the analysis of the previous section, we now assume the tunnel is evacuated of all air. (We do not consider the practical possiblity of keeping the tunnel evacuated, nor do we consider energy expenditure needed for this enormous task.) To introduce friction in the tunnel, we assume a constant frictional force $F$  by way of contact with the walls of the tunnel or, say, between the transport vehicle and the rails upon which it travels. Using this constant contact fricitional force, we now consider three different physical senarios describing the density of the Earth.

\subsection{Uniform Density}
As in the introduction, assume the Earth has constant density $\rho_0$, which implies that the gravitational field within the Earth is linear as a function of radial position. The resulting equation of motion will be the differential equation \eqref{osc_nofric}, but because we have a tunnel with constant contact friction, the equation is modified by introducing a frictional term that depends on the direction of the instantaneous velocity:
\begin{equation}
\ddot{r}+\omega^2 r = -\text{sgn}(\dot{r})f, \label{osc_constdens_constfric}
\end{equation}
where sgn is the sign (or signum) function and 
\begin{equation}
\omega^2=\frac{4\pi G\rho_0}{3}=\frac{g}{R} \;\;\; \text{and} \;\;\; f=\frac{F}{m}. \label{omega_constfric}
\end{equation}
Note that $f$ is just the frictional force per unit mass on the transport vehicle, which, for efficiency purposes from Section IIB above, we are assuming to be $f=0.35$ N/kg. The general solution is found by joining the solution to equation \eqref{osc_nofric}, which is just simple harmonic motion, with a constant:
\begin{equation}
r(t)=A\cos(\omega t - \phi) - \text{sgn}(\dot{r}) B, \label{position_constdens_constfric}
\end{equation}
such that the constant term is given by
\begin{equation}
B=\frac{f}{\omega^2}=\frac{3f}{4\pi G\rho_0}=\frac{fR}{g}, \label{constint_constdens_constfric}
\end{equation}

Though this is the \emph{mathematical} solution, a little thought about the problem shows the \emph{physical} solution to be more complicated \cite{constantfric}. Physically, the value of $A$ cannot be constant throughout the whole motion; otherwise the motion would exhibit simple harmonic motion of the same amplitude forever and the transport vehicle would never slow down. Mathematically, if $A$ were constant, the position in equation \eqref{position_constdens_constfric} would be discontinuous at each turning point. So equation \eqref{position_constdens_constfric} really represents a family of solutions, and from the form of the solution in \eqref{position_constdens_constfric}, the motion is simple harmonic between turning points (within a half-cycle or at a half period of oscillation) such that the amplitude decreases between turning points. Note the turning points occur at regularly spaced intervals in time, given by
\begin{equation}
\Delta t=\frac{T}{2}=\frac{\pi}{\omega}, \label{deltat_constdens_constfric}
\end{equation}
where $T$ is the period of oscillation, and which occur with a frequency of 
\begin{equation}
\Delta\omega = \frac{2\pi}{\Delta t}=2\omega. \label{deltaomega_constdens_constfric}
\end{equation}
In fact, we will show the amplitude of the motion $A$ decreases in a \emph{linear} fashion between turning points of the motion.

Assume the transport vehicle begins from rest at the surface of the Earth at $t=0$, producing $\phi =0$ in equation \eqref{position_constdens_constfric}. The first turning point occurs at $t_1=\pi/\omega$ and subsequently the $n^{th}$ turning point occurs at $t_n=n(\pi/\omega)$. During the half-cycle between $t_n$ and $t_{n+1}$, the motion is given by
\begin{equation}
r(t)=A_n \cos(\omega t) + (-1)^n B. \label{position_btwn_n_constdens_constfric}
\end{equation}
At the start of the half-cycle, the amplitude is $A_n$ and, at the end of the half-cycle, it decreases to $A_{n+1}$. To find the relationship between these values, we could demand that $r(t)$ be a continuous function and enforce continuity conditions at the turning points. Instead, we will derive it using the conservation of energy:
\begin{equation}
\frac{1}{2}m\omega^2 A_n^2=\frac{1}{2}m\omega^2 A_{n+1}^2 + F(A_n+A_{n+1}), \label{consenergy_constdens_constfric}
\end{equation}
from which we find
\begin{equation}
A_n^2 - A_{n+1}^2=2B(A_n + A_{n+1}), \label{recurrenceA_constdens_constfric}
\end{equation}
or
\begin{equation}
A_{n+1} = A_n - 2B. \label{diff_btwn_A_constdens_constfric}
\end{equation}
Therefore the amplitude decreases linearly by an amount $2B$ per half-cycle, from which we can conclude
\begin{equation}
A_n=A_0-2nB. \label{An_constdens_constfric}
\end{equation}
Plotting this motion in Mathematica results in Figure \ref{Const Fric Const Density 1} below.

\begin{figure}[h!]
\includegraphics[width=140mm]{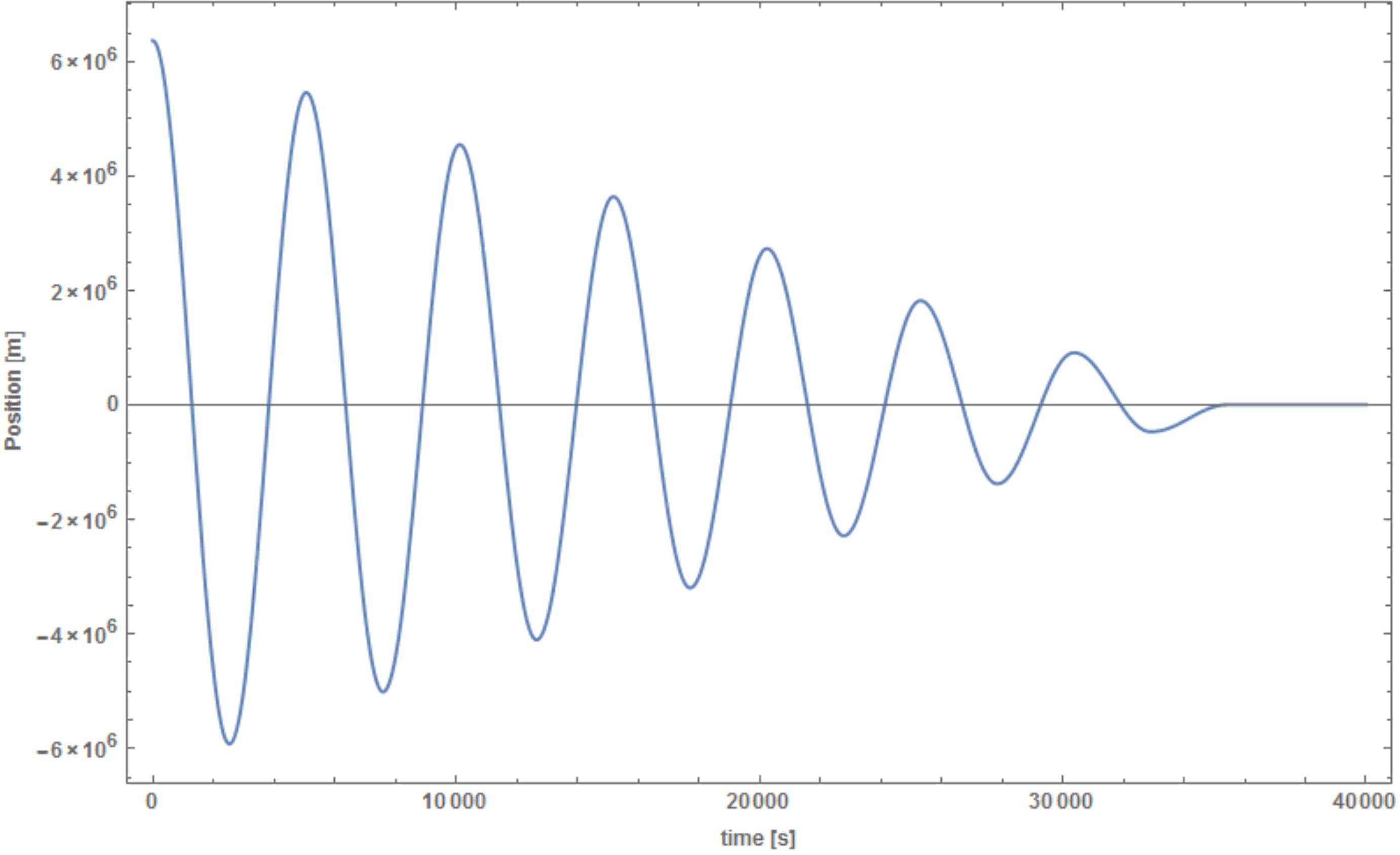}
\caption{Motion with constant density profile and constant contact friction.}
\label{Const Fric Const Density 1}
\end{figure}

This relationship, in turn, enables us to predict exactly when the motion will come to a halt. The slope of the enveloping line of the motion is given by
\begin{equation}
\frac{A_n-A_{n+1}}{\pi/\omega [n-(n+1)]} = \frac{A_n-(A_n - 2B)}{-\pi/\omega}=-\frac{2B\omega}{\pi} \label{slope_env_line_constdens_constfric}
\end{equation}

\begin{figure}[h!]
\includegraphics[width=140mm]{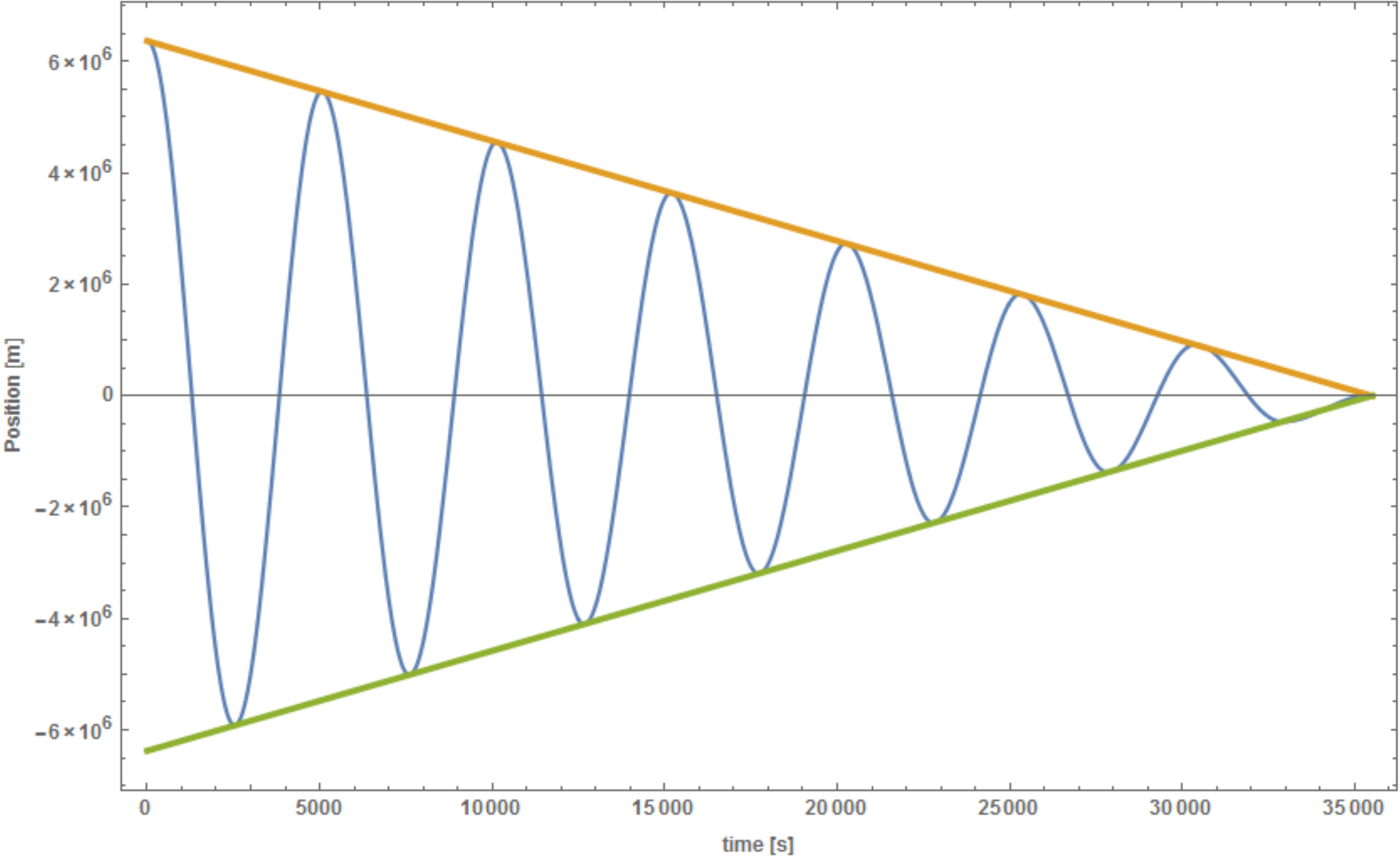}
\caption{Envelope of the motion under constant density and constant contact friction.}
\label{UniformDensEnv}
\end{figure}

and since $r(0)=R$, where $R$ is the radius of the Earth and the starting point of the transport vehicle, 
the equation of the enveloping line to the curve is given by
\begin{equation}
r_{env}(t)=-\frac{2B\omega}{\pi}t+R \label{env_line_constdens_constfric}
\end{equation}
We therefore conclude that after a time of 
\begin{equation}
t_{halt}=\frac{\pi R}{2B\omega} \label{halting_time_constdens_constfric}
\end{equation}
the transport vehicle will halt, and its position will be at the center of the Earth, $r(t_{halt})=0$. From this, we can also determine the number of complete oscillations the transport vehicle makes about the center of the Earth. Since the $n^{th}$ turning point occurs at time $t_n=n(\pi/\omega)$, we can equate this and equation \eqref{halting_time_constdens_constfric} to arrive at
\begin{equation}
n_{halt}=\left\lfloor \frac{R}{2B} \right\rfloor, \label{half_oscillations_constdens_constfric}
\end{equation}
where $\left\lfloor x \right\rfloor$ is the floor function (or greatest integer function), defined as the largest integer less than or equal to $x$. Equation \eqref{half_oscillations_constdens_constfric} gives the number of half-cycles of the motion, and, if we incorporate equation \eqref{constint_constdens_constfric}, we find the number of complete oscillations is given by
\begin{equation}
\frac{n_{halt}}{2}=\left\lfloor \frac{R}{4B} \right\rfloor =\left\lfloor \frac{g}{4f}\right\rfloor \label{total_oscillations_constdens_constfric}
\end{equation}

It is an interesting exercise to compute the total distance the transport ship travels as it oscillates about the center of the Earth, if nothing interfered with its motion. Using equations\eqref{constint_constdens_constfric}, \eqref{An_constdens_constfric}, and \eqref{half_oscillations_constdens_constfric} and the fact that $A_0=R$, the total distance, $d$, is given by
\begin{align}
 d &=A_0+ \sum_{i=1}^{n_{halt}}2A_i=A_0+\sum_{i=1}^{n_{halt}}2(A_0-2iB) \nonumber \\
    &= A_0+2n_{halt}A_0-2Bn_{halt}(n_{halt}+1) \nonumber \\
    &= R+\frac{R^2}{B}-\frac{R^2}{2B}-R=\frac{R^2}{2B} \nonumber \\
    &= \frac{gR}{2f} \label{total_dist_constdens_constfric}
\end{align}
Both equations \eqref{total_oscillations_constdens_constfric} and \eqref{total_dist_constdens_constfric} can be verified in Mathematica: the total distance the transport vehicle travels before coming to rest is 89,194,000 m according to Mathematica, and equation \eqref{total_dist_constdens_constfric} gives 89,319,000 m, which has a percent difference of 0.14\%. The difference lies in the fact that the total distance may not be an integer number of oscillations; the transport vehicle travels slightly more or less than the distance in an integer number of oscillations to come to rest at the core. 

\subsection{Constant Gravitational Acceleration}
A recent paper in this journal \cite{klotz} by Klotz examined a simpler approximation than uniform density by assuming the density profile of the Earth allows for a constant magnitude gravitational force throughout the interior of the Earth. In this case, the gravitational acceleration is always 9.8 $\text{m/s}^2$, pointing to the center of the Earth. As Klotz admits, this is a rather unphysical assumption, but it results in a travel tunnel time (38 minutes, 0 seconds) that much better approximates the more realistic travel tunnel time using the PREM density profile (38 minutes, 11 seconds), compared to the tunnel travel time using the uniform density profile (42 minutes, 12 seconds). This analysis was done, however, without considering any type of frictional forces in the tunnel. When constant contact friction is introduced, the motion dynamics of the tunnel transport vehicle using a density profile that gives rise to a constant gravitational acceleration still better approximates the motion using the PREM density profile compared to the uniform density model.

The equation of motion in the case of constant gravitational acceleration $g$ with a constant frictional force per unit mass of the transport vehicle of $f$ is easily deduced to be
\begin{equation}
\ddot{r}=-g\;\text{sgn}(r)-f\text{sgn}(\dot{r}), \label{diffeqn_const_g_const_f}
\end{equation}
with the initial values $\dot{r}(0)=0$ and $r(0)=R$, since we are assuming the transport vehicle is released from rest at the surface of the Earth ($r=R$). Upon direct integration, the position as a function of time is found to be
\begin{equation}
r(t)=-\frac{gt^2}{2}\text{sgn}(r)-\frac{ft^2}{2}\text{sgn}(\dot{r}) + R. \label{position_const_g_const_f}
\end{equation}
Due to the direction of the gravitational acceleration and the frictional force, this one equation actually represents four different equations per oscillation of the motion. Breaking this down by which region of the tunnel the transport vehicle is in, we have
\begin{align}
r_0(t) &= -\frac{g}{2}t^2+\frac{f}{2}t^2+R, \;\;\; 0 \leq t \leq t_0 \label{1st}\\
r_1(t) &= \frac{g}{2}t^2+\frac{f}{2}t^2 + c_1(t), \;\;\; t_0 \leq t \leq t_1 \label{2nd}\\
r_2(t) &= \frac{g}{2}t^2-\frac{f}{2}t^2+c_2(t), \;\;\; t_1 \leq t \leq t_2 \label{3rd}\\
r_3(t) &= -\frac{g}{2}t^2-\frac{f}{2}t^2 +c_3(t), \;\;\; t_2 \leq t \leq t_3 \label{4th}
\end{align}
where $c_1$, $c_2$, and $c_3$ are constants of integration and can depend on time. These individual equations represent the position of the transport vehicle as a function of time in four different zones. Equation \eqref{1st} describes the motion of the vehicle as it falls from the surface of the Earth toward the center, equation \eqref{2nd} is the motion after it passes the center and approaches, but doesn't reach, the antipodal of its starting position, equation \eqref{3rd} is the motion of the vehicle as it falls back toward the center of the Earth in the opposite direction, and finally equation \eqref{4th} represents the motion as it passes the center of Earth again, approaching, but doesn't reach, its original starting position. These equations only represent the first complete oscillation of the transport vehicle, and they are related by continuity conditions at the center of the Earth given by
\begin{align}
r_0(t_0) = r_1(t_0), \;\; &\text{and} \;\; \dot{r}_0(t_0)=\dot{r}_1(t_0),\\
r_2(t_2) = r_3(t_2), \;\; &\text{and} \;\; \dot{r}_2(t_2)=\dot{r}_3(t_2),
\end{align}
and at the turning point, 
\begin{equation}
r_1(t_1) = r_2(t_1), \;\; \text{and} \;\; \dot{r}_1(t_1)=\dot{r}_2(t_1).
\end{equation}
Enforcing continuity generates the initial conditions for each zone (save $r_0(t)$ in equation \eqref{1st}), thus uniquely determining the constants of integration $c_1$, $c_2$, and $c_3$. In Figure \ref{earth_tunnel_4} below, zone I corresponds to the position represented by equation \eqref{1st}, zone II to equation \eqref{2nd}, zone III to equation \eqref{3rd}, and zone IV to equation \eqref{4th}, and all together these represent one complete oscillation in the tunnel.

\begin{figure}[h!]
\includegraphics[width=70mm]{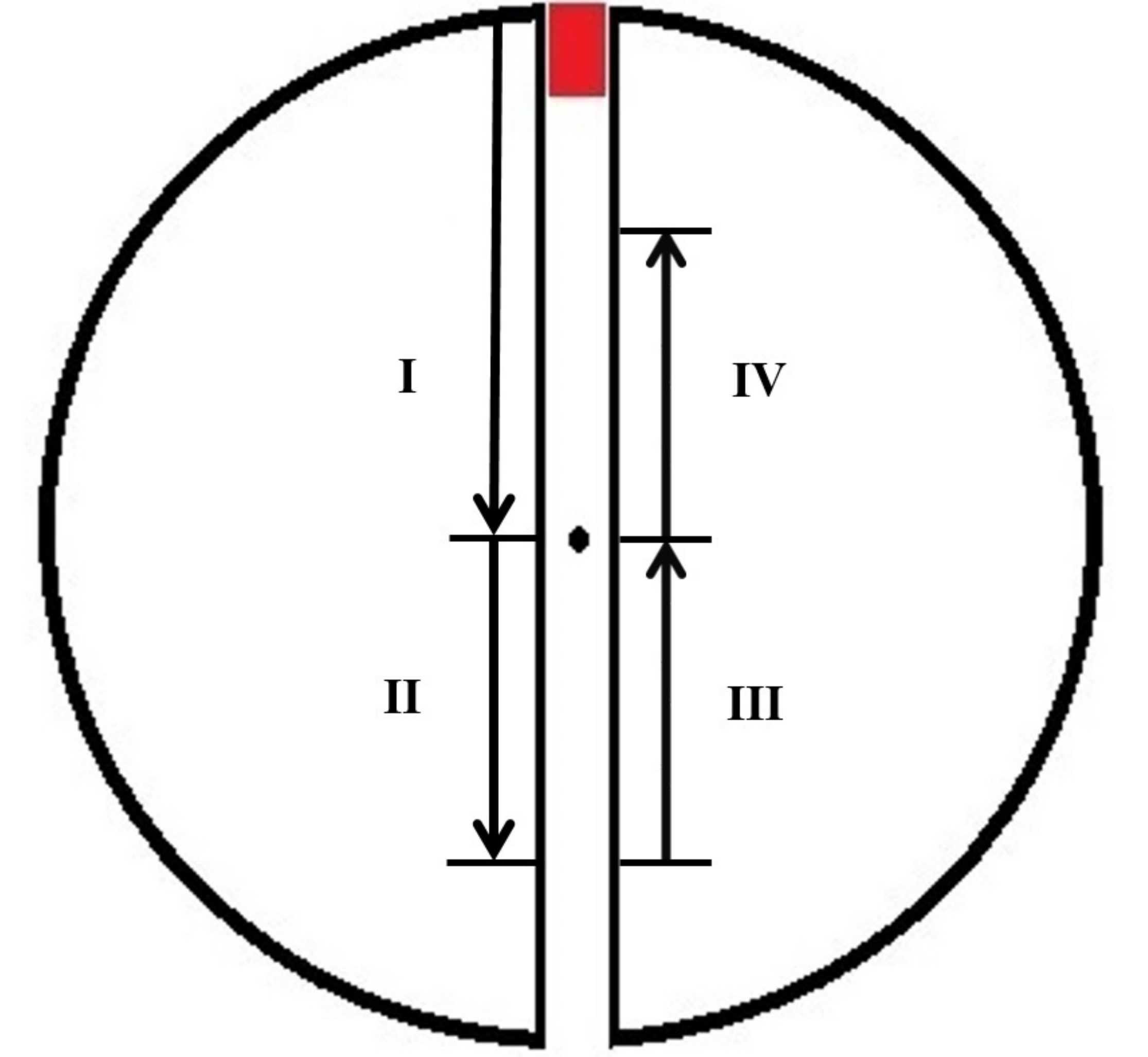}
\caption{A complete oscillation through the tunnel by zones}
\label{earth_tunnel_4}
\end{figure}

\begin{figure}[h!]
\includegraphics[width=140mm]{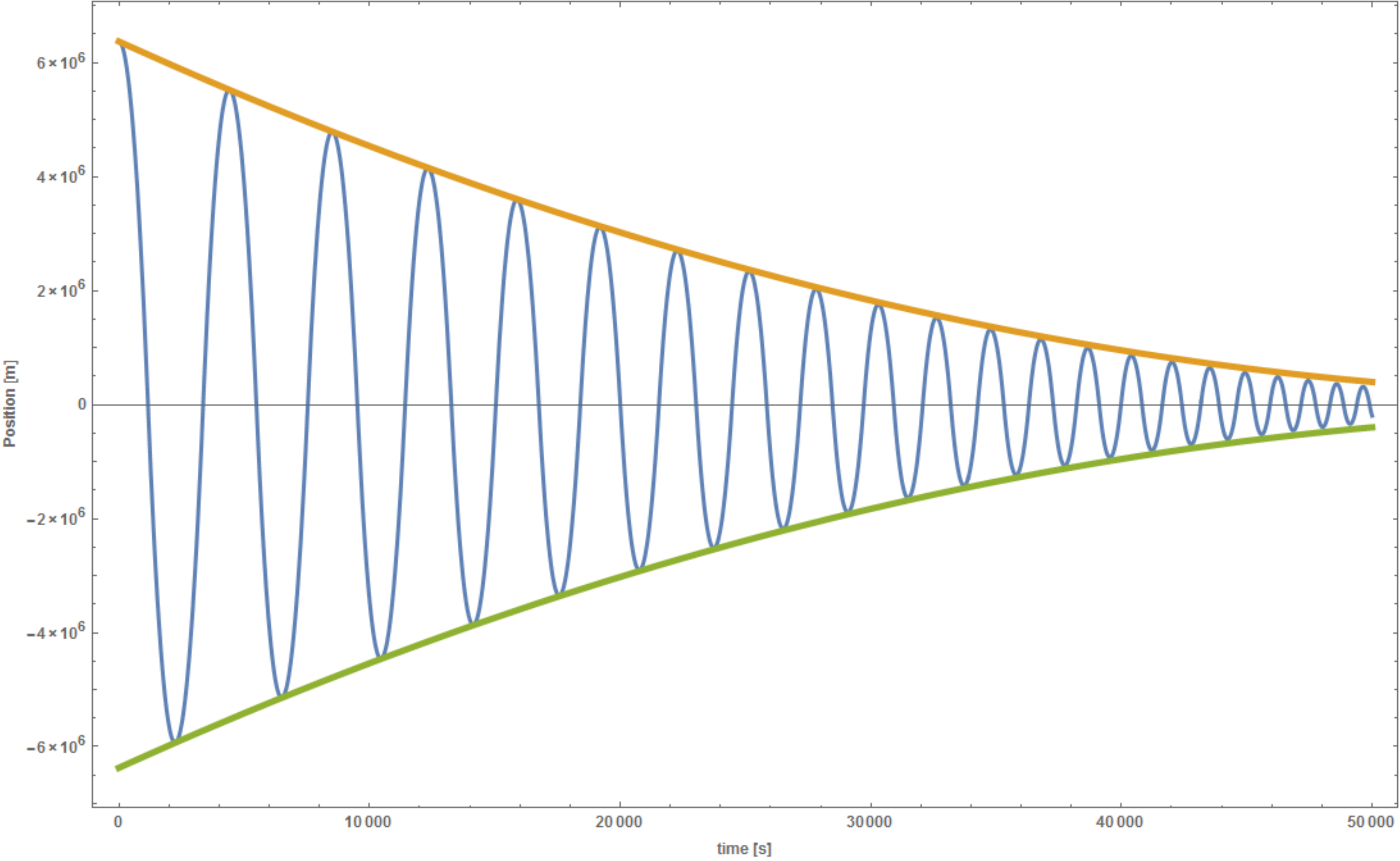}
\caption{Position as a function of time using constant $g$.}
\label{DepthConstantgEnv}
\end{figure}

Moreover, all subsequent oscillations can be described by equations of the same form as \eqref{1st} -- \eqref{4th}, but each will have different constants of integration based on the continuity conditions. For simplicity, let $2\alpha=g-f$ and let $2\beta=g+f$. Then the complete motion may be described by the equations
\begin{align}
r_i(t) &= -\alpha t^2+c_i(t), \;\;\; t_{i-1} \leq t \leq t_i \label{1st_i}\\
r_{i+1}(t) &= \beta t^2 + c_{i+1}(t), \;\;\; t_i \leq t \leq t_{i+1} \label{2nd_i}\\
r_{i+2}(t) &= \alpha t^2+c_{i+2}(t), \;\;\; t_{i+1} \leq t \leq t_{i+2} \label{3rd_i}\\
r_{i+3}(t) &= -\beta t^2 +c_{i+3}(t), \;\;\; t_{i+2} \leq t \leq t_{i+3} \label{4th_i}
\end{align}
along with the continuity conditions
\begin{align}
r_i(t_i) = r_{i+1}(t_i), \;\; &\text{and} \;\; \dot{r}_i(t_i)=\dot{r}_{i+1}(t_i),\\
r_{i+2}(t_{i+2}) = r_{i+3}(t_{i+2}), \;\; &\text{and} \;\; \dot{r}_{i+2}(t_{i+2})=\dot{r}_{i+3}(t_{i+2}),\\
r_{i+1}(t_{i+1}) = r_{i+2}(t_{i+1}), \;\; &\text{and} \;\; \dot{r}_{i+1}(t_{i+1})=\dot{r}_{i+2}(t_{i+1}),\\
r_{i+3}(t_{i+3}) = r_{i+4}(t_{i+3}), \;\; &\text{and} \;\; \dot{r}_{i+3}(t_{i+3})=\dot{r}_{i+4}(t_{i+3}).
\end{align}
for $0 \leq i \leq n$, where $n$ is some value representing the last complete oscillation, $t_{-1} \myeq 0$, and $t_i+t_{i+1}+t_{i+2}+t_{i+3}$ is the time for one complete oscillation. As evidenced by these equations, the motion is \emph{piecewise periodic}, and difficult to solve analytically. Numerically, however, we can graph this motion in Mathematica again, resulting in Figure \ref{DepthConstantgEnv}.

\subsection{Preliminary Reference Earth Model}
The PREM data \cite{prem} reconstructs the density profile of the Earth from sesimic data. The mass is now a function of $r$ and given by
\begin{equation}
M(r)=4\pi\int_0^r \rho(\xi)\xi^2d\xi, \label{mass_fcn_PREM}
\end{equation}
so that the differential equation governing the motion is
\begin{equation}
\ddot{r}-\frac{G}{r^2}M(r)= \ddot{r}-\frac{4\pi G}{r^2}\int_0^r \rho(\xi)\xi^2d\xi=-\text{sgn}(\dot{r})f. \label{osc_PREM}
\end{equation}

\begin{figure}[h!]
\includegraphics[width=140mm]{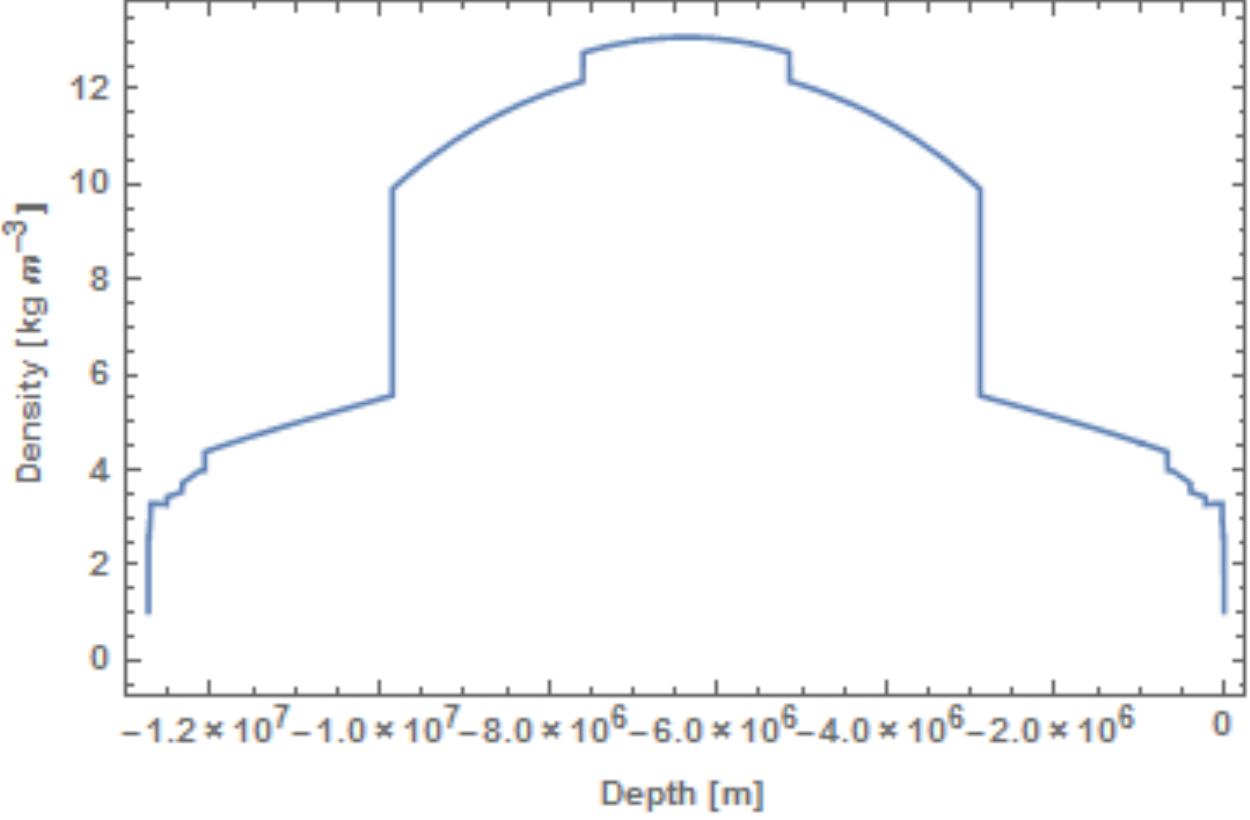}
\caption{Preliminary Reference Earth Model Data.}
\label{PREM}
\end{figure}

As the motion of the transport vehicle in the tunnel under the PREM density profile in equation \eqref{osc_PREM} cannot be solved analytically due to the piecewise continuous nature of the PREM density data, we once again utilize the power of Mathematica to instead numerically calculate the position (depth) as a function of time. The graph of the motion using the PREM data is given in Figure \ref{DepthPREMenv} below.

\begin{figure}[h!]
\includegraphics[width=140mm]{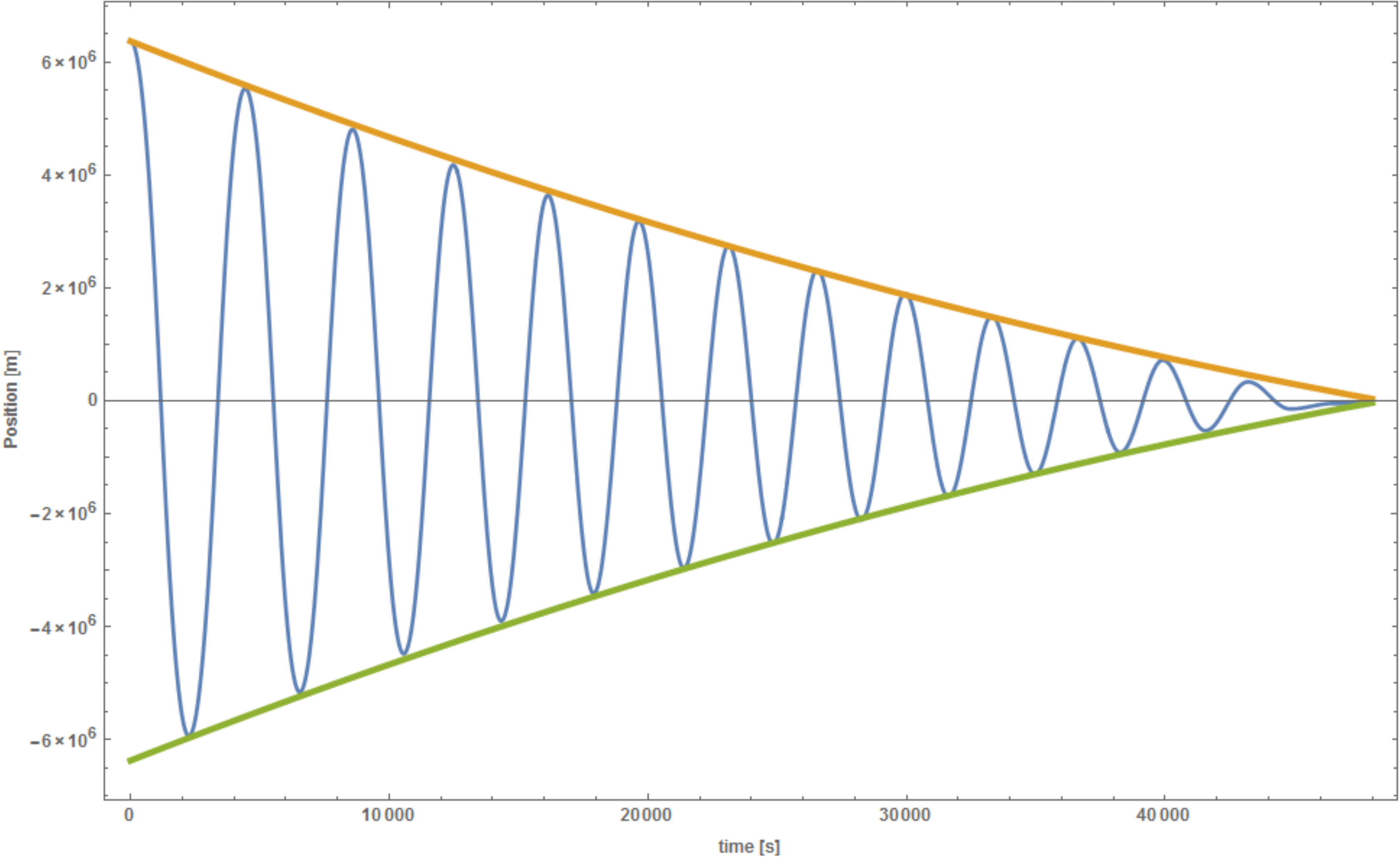}
\caption{Position as a function of time using PREM data.}
\label{DepthPREMenv}
\end{figure}

\subsection{Comparison of the Motion In the Three Models}

\begin{figure}[h!]
\includegraphics[width=140mm]{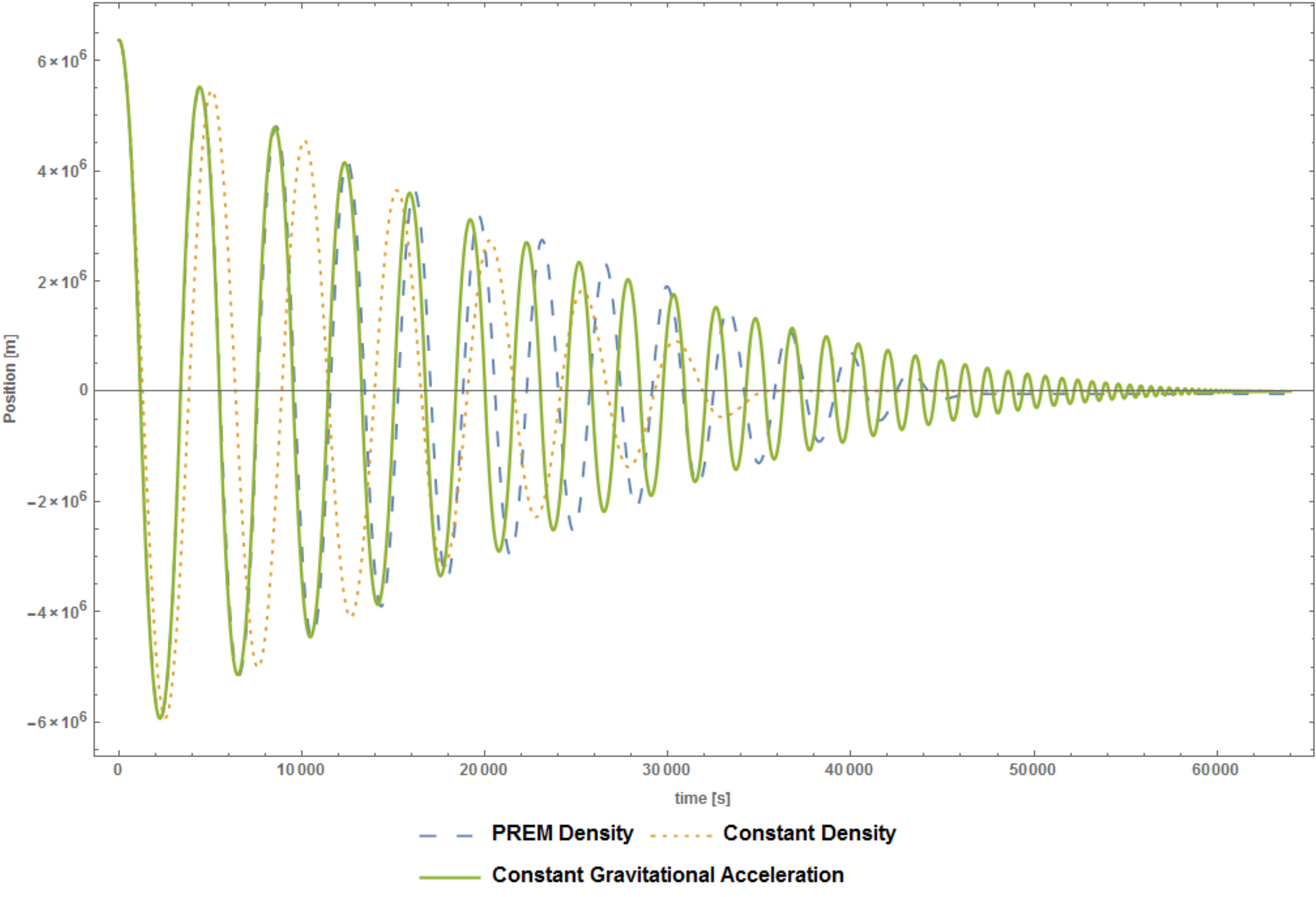}
\caption{A comparison between the models of the position as a function of time.}
\label{PosCompare}
\end{figure}

We noticed several interesting features when comparing the transport vehicle motion between the three models used above. From the graph of depth (position) as a function of time, given in Figure \ref{PosCompare}, note that the motion under the density profile corresponding to a constant gravitational acceleration better corresponds to the motion under the PREM density profile since the oscillations don't significantly deviate from one another until about 20,000 seconds into the oscillation. On the other hand, the motion under the assumption of constant density appreciably deviates from the motion under the PREM density profile after only a half-cycle of about 4000 seconds. In fact, we compared the time it takes for the transport vehicle to reach the center of the Earth under each model, without any friction and with constant contact friction for our value of $f=0.35$ in Table \ref{times_to_fall_to_center}. Note the greatest percent difference between the times with and without friction was for the assumption of a constant density Earth. More striking, however, is the percent difference between the times with and without friction for the density profile corresponding to a constant gravitational acceleration throughout the tunnel is the same as the percent difference between the times with and without friction for the more realistic PREM density profile. In addition, the time to arrive at the center of the Earth was not that much different for the constant $g$ model compared to the PREM data model, but the constant $\rho$ model differed significantly from either. This leads us to the conclusion that, although the motion under the assumption of a density profile corresponding to a constant gravitational acceleration may be unrealistic, it more closely resembles the motion under the PREM density profile, much better than the motion under the assumption of constant density with or without frictional forces present. Our analysis therefore further supports Klotz's model of a density profile corresponding to constant gravitational acceleration as a better approximation of the gravity train motion than the motion under the assumption of a uniform density Earth. 

\begin{figure}[h!]
\includegraphics[width=140mm]{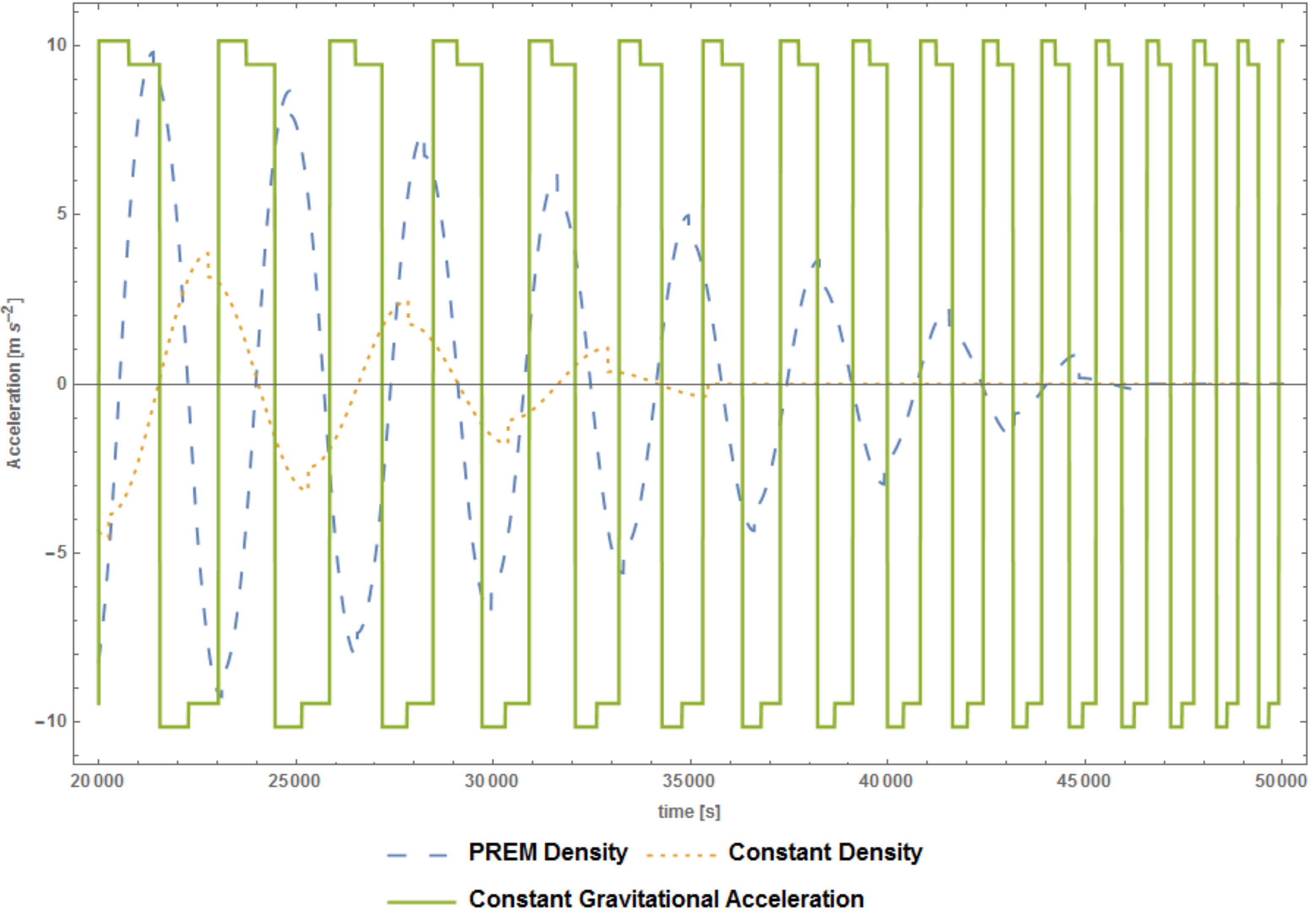}
\caption{A comparision between the models of the acceleration as a function of time.}
\label{AccelCompare}
\end{figure}

In addition, if we examine the graph comparing the accelerations of the three different models, the first features that are noticeable are the discontinuous steps that appear at the ends of each oscillation. These correspond to the sudden changes in direction of the frictional forces that work to oppose the motion of the vehicle. When the direction the the velocity changes so does the direction of these frictional forces. The other feature that is notable, particularly at the later times displayed in Figure \ref{AccelCompare}, is that both the PREM and constant gravational acceleration models achieve larger accelerations than the constant density model, but for shorter lengths of time. This results in initially remarkably similar travel times but also allows friction to affect each model differently. This then results in greater increases in the travel time for the constant density model due to the fricitonal forces and a shorter core stopping time than for the other two models.

\begin{table}[ht]
\caption{Time in Seconds to Arrive at Center of Earth}
\centering
\begin{tabular}{c c c c}
\hline\hline
Model & No Friction & Friction & \% difference \\ [0.5ex]
\hline
PREM & 1145.88 & 1167.17 & 1.84 \\ 
Constant $g$ & 1140.26 & 1161.19 & 1.82 \\
Constant $\rho$ & 1265.63 & 1295.43 & 2.33 \\ [1ex] 
\hline
\end{tabular}
\label{times_to_fall_to_center}
\end{table}

\section{Conclusions}

Building a gravity train tunnel through the Earth would not only be prohibitively expensive, it would also be quite technically challenging. As we have seen, if the tunnel is not evacuated of all air, the quadratic drag force of the air on the transport vehicle would effectively make it impossible to use as a viable means of transportation. Then we examined the motion under a constant sidewall friction between the transport vehicle and the tunnel, using three different density profile models. Though the assumption of a uniform density Earth was the most amenable to solving analytically (with or without the constant frictional force present) we saw that, with a constant frictional force, the assumption of a constant gravitational acceleration throughout the interior of the Earth most closely approximated the more realistic motion under the PREM density profile, lending further weight to a few of the conclusions in the paper by Klotz\cite{klotz}.

We also found the value of the constant frictional force used throughout Section III by considering energy expenditures, compared to a meridional trip halfway around the Earth.  But, given that the energy expenditure through the gravity tunnel with our value of the constant sidewall friction is the same as that of an Airbus A380 meridonal trip halfway around the Earth, we concluded that since the travel time between antipodal points is far less through the tunnel than it would be for a meridonal trip, future societies may very well benefit from building an evacuated tunnel between key antipodal points on the Earth.

\begin{acknowledgments}

We gratefully acknowledge Artur Tsobanjan and his invaluable help with Mathematica. 

\end{acknowledgments}

\end{document}